\documentclass[
    ,final            
    ,numberedheadings 
    ,cmfonts          
  ]
  {aipproc}

\layoutstyle{6x9}
\newcommand{\beq}{\begin{eqnarray}}
\newcommand{\eeq}{\end{eqnarray}}

\begin{document}

\title{Pion-Photon Transition Distribution Amplitudes}


\author{A. Courtoy }{
  address={Departamento de F\'{\i}sica Te\'orica and Instituto de F\'{\i}sica Corpuscular,\\
Universidad de Valencia-CSIC,\\ E-46100 Burjassot (Valencia), Spain.\\},
altaddress={ E-mail: aurore.courtoy@uv.es}
}

\author{S. Noguera}{
  address={Departamento de F\'{\i}sica Te\'orica and Instituto de F\'{\i}sica Corpuscular,\\
Universidad de Valencia-CSIC,\\ E-46100 Burjassot (Valencia), Spain.\\},
altaddress={ E-mail: santiago.noguera@uv.es}
}

\begin{abstract}
  The newly introduced Transition Distribution Amplitudes (TDAs) are
  discussed for the $\pi$-$\gamma$ transitions.  Relations between $\pi$-$\gamma$  and $\gamma$-$\pi$ TDAs for different cases are given. Numerical values for
  the $\pi$-$\gamma$ TDAs in different models are compared.
 GPD's features are extended to TDAs and the role of PCAC
  highlighted. We give hints for the evaluation of cross sections for
  meson pair production in our approach.
\end{abstract}

\maketitle

\section{Introduction}

\begin{figure}
\includegraphics[height=.18\textheight]{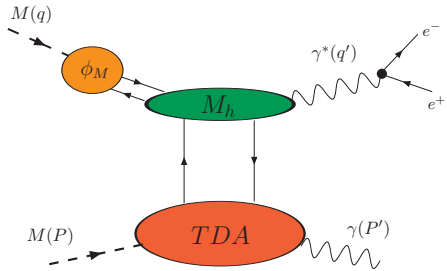}
  \caption{Factorization of the amplitude for the process \eqref{hh}. }
\label{facto}
\end{figure}

One of the  main open questions in particle physics is the
understanding of the structure of hadrons in terms of quarks.  An
important tool for such a purpose is provided by hard processes. The
large virtuality $Q^2$ involved in these processes allows the
factorization of their amplitudes into hard and soft contributions.
The  hard  contribution to the scattering amplitude is known from
perturbative QCD but the interesting quantity unveiling the structure
of hadrons is the soft or nonperturbative contribution. In Deep
Inelastic Scattering (DIS) this nonperturbative quantity is expressed
in terms of the parton distribution functions (PDF).  PDFs can be
expressed as the Fourier transform of a bilocal current between the
same initial and final hadronic state.  Generalized parton
distributions (GPD) \cite{Radyushkin:1996nd, Ji:1996ek, Diehl:2003ny}
extend this concept to off-diagonal matrix elements of the same
currents. GPD govern exclusive processes with the same hadron in the
initial and final state in the soft part of the process but with
momentum transfer different from zero. Deeply virtual
Compton scattering (DVCS) is a tipical example of processes governed by GPDs. Recently a further ``generalization'' to
transition distribution amplitudes (TDA) in which the initial and the
final state in the soft part of the amplitude are different has been
introduced. They have been defined for processes like hadron
annihilation into two photons and backward VCS in the kinematical
regime where the virtual photon is highly off-shell but with small
momentum transfer $t$ \cite{Pire:2004ie} \beq H\bar H\to
\gamma^{\ast}\gamma \to e^+ e^-\gamma \quad
\mbox{and}\quad\gamma^{\ast} H\to H\gamma\quad,
\label{hh}
\eeq with $H$ a hadron, or exclusive meson pair production in
$\gamma^{\ast}\gamma$ scattering in the same kinematical regime
\cite{Lansberg:2006fv}
 \beq \gamma_L^{\ast}\gamma \to
M^{\pm}\pi^{\mp}\quad,
\label{gg}
\eeq
 $M$ being either $\rho_L$ or $\pi$.  Advocating
\cite{Pire:2004ie} that the factorization theorems for exclusive
processes can be extended to the processes under consideration, i.e.
\eqref{hh} and \eqref{gg}, the amplitude can be factorized as shown in
Fig.~\ref{facto} with the TDAs, describing the $\pi\to\gamma$
transition, being the Fourier transform of the matrix element of bilocal currents
at a light-like distance.

The nonperturbative nature of the distribution functions imposes the
use of effective theories, models or phenomenological
parametrizations.  In Ref.~\cite{Courtoy:2007vy} the calculation of
the $\pi$-$\gamma$ TDAs in a covariant Bethe-Salpeter approach has
been defined. All the invariances of the problem are hence preserved,
e.g. gauge and translational invariance. For the numerical evaluation, the
Nambu - Jona-Lasinio (NJL) model for the description of the pion has
been used. The Pauli-Villars
regularization scheme has been chosen because it is Lorentz invariant.

  Estimates for the $\pi$-$\gamma$ TDAs have   been given
in Ref.~\cite{Tiburzi:2005nj} and  a calculation has been performed in the Spectral Quark
Model (SQM)~\cite{Broniowski:2007fs}. Both studies parametrize the
TDAs by means of double distributions \cite{Radyushkin:1996nd}.  A detailed comparison of the SQM and the NJL models in the determination of the pion GPD can be found in Ref.~\cite{Broniowski:2007si}. Finally in
Ref.~\cite{Kotko:2008gy} TDAs have been calculated in a non-local
chiral quark model, confirming the results of the previous
calculations.

In the following Section we introduce the vector and axial TDAs
emphasizing their connection to the pion transition form factors,
appearing in the radiative pion decay \cite{Moreno:1977kx}. The
presence of an additional contribution due to PCAC is explicitly
shown.  Numerical results and a comparison between the different
models are presented in Section~\ref{sec:3}.

\section{Definition of the Pion-Photon TDAs}
\label{sec:2}

General arguments, such as Lorentz invariance, lead to some important properties of GPDs.
Taking their first Mellin moment, one can relate GPDs to the
corresponding form factors through the sum rules.  Also their higher
Mellin moments are polynomials in the skewness variable $\xi$ by
Lorentz invariance.

\begin{figure}
\includegraphics[height=.1\textheight]{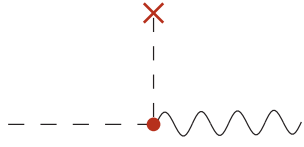}
  \caption{Pion pole contribution between the axial current (the cross) and the incoming pion-photon vertex.}
\label{pipo}
\end{figure}

Similarly, as a consequence of Lorentz invariance, TDAs are
constrained by sum rules and polynomial expansions. The first Mellin moments of 
$\pi$-$\gamma$ transition distribution functions are  related
to the vector and axial-vector transition form factors, $F_V$ and
$F_A$, through the sum rules. The definition of these form factors is
given from the vector and axial-vector currents \cite{Moreno:1977kx}
\beq \left\langle \gamma(p') \right\vert \bar{q}(0)
\gamma_{\mu}\tau^{-}q( 0) \left\vert \pi(p) \right\rangle
&=&-i\,e\,\varepsilon^{\nu}\,\epsilon_{\mu\nu\rho\sigma}\,p^{'\rho}\,p^{\sigma}\,\frac{F_{V}(t)
}{m_{\pi}}\quad,
\label{ffv}\\
\left\langle \gamma(p') \right\vert \bar{q}(0)
\gamma_{\mu}\gamma_{5}\tau^{-}q(0) \left\vert\pi(p) \right\rangle &
=&e\,\varepsilon^{\nu}\left( p_{\mu
  }^{\prime}\,p_{\nu}-g_{\mu\nu}\,p^{\prime}.p\right)
\frac{F_{A}\left(
    t\right)  }{m_{\pi}}\nonumber\\
&& +e\,\varepsilon^{\nu}\left( \left( p^{\prime}-p\right)
  _{\mu}\,p_{\nu
  }\frac{2\sqrt{2}f_{\pi}}{m_{\pi}^{2}-t}-\sqrt{2}f_{\pi}\,g_{\mu\nu}\right)
\quad,
\label{ff}%
\eeq with $f_{\pi}=92.4$ MeV, $\varepsilon^{0123}=1$ and
$\tau^{-}=\left( \tau_{1}-i\,\tau_{2}\right) /2$.  All the structure
of the decaying pion is included in the form factors $F_{V}$ and
$F_{A}$. The vector current only contains a Lorentz structure
associated with the $F_{V}$ form factor.  The axial form factor $F_A$
also gives the structure of the pion but contains additional terms
required by electromagnetic gauge invariance. The second term on the
right-hand side of Eq.~\eqref{ff}, which corresponds to the axial
current for a point-like pion, contains a pion pole coming from the
pion inner bremsstrahlung: the incoming pion and outgoing photon
couple with the axial current through a virtual pion (Fig.~\ref{pipo})
as required by the Partial Conservation of the Axial Current (PCAC).
The third term in Eq.~\eqref{ff} is a pion-photon-axial current
contact term, proportional to $f_{\pi}g_{\mu\nu}$.

Going to the TDAs we introduce the light-cone coordinates
$v^{\pm}=\left( v^{0}\pm v^{3}\right) /\sqrt{2}$ and the transverse
components $\vec{v}^{\bot}=\left( v^{1},v^{2}\right) $ for any
$4$-vector $v^{\mu}$. We define $P=\left( p+p^{\prime}\right) /2$ and
the momentum transfer, $\Delta=p^{\prime}-p,$ therefore
$P^{2}=m_{\pi}^{2}/2-t/4$ and $t=\Delta^{2}$. The skewness variable
describes the loss of  momentum in the light front direction of the incident pion,
i.e. $\xi=\left( p-p^{\prime}\right) ^{+}/2P^{+}$.  Its value ranges
between $t/\left( 2m_{\pi}^{2}-t\right) <\xi<1$. Negative values of
the skewness variable can be allowed.  Regarding the real photon
polarization, $\varepsilon^{\nu}$, we have the transverse condition
$\varepsilon.p'=0$ and an additional gauge fixing condition. We assume
that this condition is such that $\varepsilon^+/P^+$ is kinematically
higher twist. The standard gauge fixing conditions, $\varepsilon^0=0$
or $\varepsilon^+=0$, both satisfy the previous requirement.  To
leading twist, the TDAs are therefore defined
\beq \int\frac{dz^{-}}{2\pi}e^{ixP^{+}z^{-}}\left.
  \left\langle \gamma\left( p^{\prime}\right) \right\vert
  \bar{q}\left( -\frac{z}{2}\right) \gamma
  ^{+}\hspace*{-0.05cm}\tau^{-}q\left( \frac{z}{2}\right) \left\vert
    \pi^+\left( p\right)
  \right\rangle \right\vert _{z^{+}=z^{\bot}=0} \nonumber\\\nonumber\\
\hspace{2cm}=i\,e\,\varepsilon_{\nu}\,\epsilon^{+\nu\rho\sigma}\,P_{\rho}\,\Delta_{\sigma
}\,\frac{V^{\pi^{+}\to\gamma}\left(  x,\xi,t\right)  }{\sqrt{2}f_{\pi}}\nonumber\quad,\\
\nonumber\\
\int\frac{dz^{-}}{2\pi}e^{ixP^{+}z^{-}}\left.  \left\langle
    \gamma\left( p^{\prime}\right) \right\vert \bar{q}\left(
    -\frac{z}{2}\right) \gamma
  ^{+}\hspace*{-0.05cm}\gamma_{5}\tau^{-}q\left( \frac{z}{2}\right)
  \left\vert \pi^+\left( p\right) \right\rangle \right\vert
_{z^{+}=z^{\bot}=0} \nonumber\\\nonumber\\
=e\,\left( \vec{\varepsilon}^{\bot}\cdot\vec{\Delta}^{\bot}\right)
\frac{A^{\pi^{+}\to\gamma}\left( x,\xi,t\right) }{\sqrt{2}f_{\pi}}
+e\,\left( \varepsilon\cdot\Delta\right)
\frac{2\sqrt{2}f_{\pi}}{m_{\pi }^{2}-t}~\epsilon\left( \xi\right)
~\phi\left( \frac{x+\xi}{2\xi}\right) \quad,
 \label{axcurr}
 \eeq with $\epsilon(\xi) $ equal to $1$ for $\xi>0,$ and equal to
 $-1$ for $\xi<0$. Here $V(x,\xi,t) $\ and $A( x,\xi,t) $ are
 respectively the vector and axial TDAs.  Hence the axial matrix element
 contains the axial TDA and the pion pole contribution that
 has been isolated in a model independent way \cite{Tiburzi:2005nj,
   Courtoy:2007vy, Pire:2007ut}.  The latter term is parametrized by a
 point-like pion propagator multiplied by the distribution amplitude
 (DA) of an on-shell pion, $\phi(x)$. Notice that the pion DA obeys
 the normalization condition $\int_0^1 dx \,\phi(x)=1$; the
 connection through the sum rules of Eq.~\eqref{axcurr} with
 Eqs.~\eqref{ffv}  and  \eqref{ff} is therefore obvious.

The contribution of a pion pole is not a new feature
 of large-distance distributions. TDAs like GPDs are low-energy
 quantities in QCD though their degrees of freedom are quarks and
 gluons.  One thus expects chiral symmetry to manifest itself, what
 implies a matching between the degrees of freedom of parton
 distributions and the low-energy degrees of freedom such as pions.
 Actually, in the region $x\in [-|\xi|, |\xi|]$, the
 emission of a $q \bar q$ pair from the initial state can be
 assimilated to a meson distribution amplitude. 

Here we have defined the TDAs in the particular case of a transition
from a $\pi^+$ to a photon, parametrizing the processes given by
Eq.~\eqref{hh}. Symmetries relate the latter distributions to TDAs
involved in other processes. For instance, we could wish to study the
$\gamma$-$\pi^-$ TDAs entering the factorized amplitude of the process
\eqref{gg}. 

Unifying our notations with the notations of
Ref.~\cite{Lansberg:2006fv}, we define the $\gamma$-$\pi^{\pm}$ TDAs
 \beq
&&\int\frac{dz^{-}}{2\pi}e^{ixP^{+}z^{-}}\left.  \left\langle \pi^{\pm}(p)
  \right\vert \bar{q}\left( -\frac{z}{2}\right) \gamma
  ^{+}\hspace*{-0.05cm}\tau^{\pm}q\left( \frac{z}{2}\right)
  \left\vert\gamma( p^{\prime}\varepsilon)
  \right\rangle \right\vert _{z^{+}=z^{\bot}=0} \nonumber\\\nonumber\\
&&=i\,e\,\varepsilon_{\nu}\,\epsilon^{+\nu\rho\sigma}\,P_{\rho}\,(p-p')_{\sigma
}\,\frac{V^{\gamma\to\pi^{\pm}}(  x,-\xi,t)  }{\sqrt{2}f_{\pi}}\nonumber\quad,\\
\nonumber\\
&&\int\frac{dz^{-}}{2\pi}e^{ixP^{+}z^{-}}\left.  \left\langle
    \pi^{\pm}\left( p\right) \right\vert \bar{q}\left( -\frac{z}{2}\right)
  \gamma^+\hspace*{-0.05cm}\gamma_{5}\tau^{\pm}q\left(
    \frac{z}{2}\right) \left\vert\gamma\left(
      p^{\prime}\varepsilon\right)   \right\rangle \right\vert _{z^{+}=z^{\bot}=0} \nonumber\\\nonumber\\
&& =-e\,\left( \vec{\varepsilon}^{\bot}\cdot(\vec{p}^{\bot
  }-\vec{p'}^{\bot})\right) \frac{A^{\gamma\to\pi^{\pm}}(x,-\xi,t)
}{\sqrt{2}f_{\pi} } \nonumber\\
&&\pm e\,\left( \varepsilon\cdot(p-p')\right)
\frac{2\sqrt {2}f_{\pi}}{m_{\pi}^{2}-t}~\epsilon(-\xi)
~\phi\left( \frac{x+\xi}{2\xi}\right) \quad,
\label{tdarev}
\eeq where we have preserved the definition $\xi=(p-p')^+/(p+p')^+$
given before the
Eq.~\eqref{axcurr}.

Time reversal relates the $\pi^+$-$\gamma$ TDAs to $\gamma$-$\pi^+$
TDAs in the following way
 \beq
D^{\pi^+\to\gamma}(x,\xi,t)&=&D^{\gamma\to\pi^+}(x,-\xi,t)\quad,
 \eeq
where $D=V,A$. And $CPT$ relates the presently calculated TDAs to
their analog for a transition from a photon to a $\pi^-$
\beq
V^{\pi^+\to\gamma}(x,\xi,t)&=&V^{\gamma\to\pi^-}(-x,-\xi,t)\quad,\nonumber\\
A^{\pi^+\to\gamma}(x,\xi,t)&=&-A^{\gamma\to\pi^-}(-x,-\xi,t)\quad.
\eeq

\section{Discussion}
\label{sec:3}

Basic properties of GPDs and TDAs, like sum rules and polynomiality
are related to Lorentz and gauge symmetries. Therefore, we need a
method of calculation which preserves these properties. The main
problem here is the description of hadrons, as bound states of quarks,
preserving these symmetries. One solution is to use a field
theroretical formalism, with a covariant Bethe-Salpeter approach for
the description of the hadrons.  In this formalism, GPDs and TDAs are
integrals over the Bethe-Salpeter amplitudes. This method has been
developed in Refs.~\cite{Courtoy:2007vy,Theussl:2002xp} for local
lagrangians and in Refs.~\cite{Noguera:2005ej, Noguera:2005cc} for
non-local lagrangians. In the case of pions, the simplest realistic
model which realizes these ideas is the NJL model within the
Pauli-Villars regularization scheme. We will therefore use this model
for the discussion of the calculation.

\begin{figure}
\centering
  \mbox{\includegraphics[height=.1\textheight]{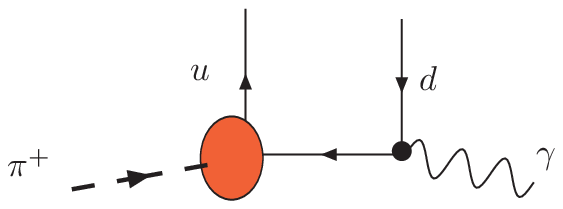}
\includegraphics[height=.2\textheight]{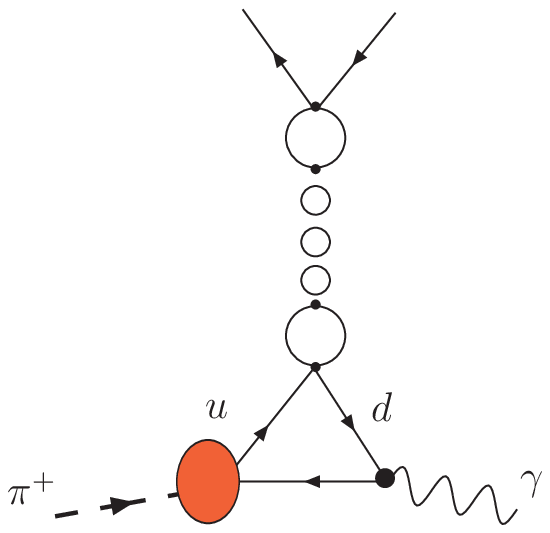}}
\caption{Diagrams contributing to pion-photon TDA, shown here for an active $u$-quark. 
The diagram on the right corresponds to a pion ``rescattering'' .}
\label{feyn}
\end{figure}

We consider that the process is dominated by the handbag diagram.
Each TDA has two related contributions \cite{Courtoy:2007vy},
depending on which quark ($u$ or $d$) of the pion is scattered off by
the deep virtual photon.  The leading contributions to the handbag
diagram are depicted in Fig. \ref{feyn} for an active $u$-quark, with
the diagram on the right corresponding to a coupling of the bilocal
current to a quark-antiquark pair. The vector TDA receives
contribution only from the first type of diagram, i.e. the diagram on the left of
Fig. \ref{feyn}. In the case of the axial TDA, a contribution in the
$-|\xi|<x<|\xi|$ region arises from the second diagram of
Fig. \ref{feyn}. This second contribution comes from the re-scattering
of a $q\bar{q}$ pair in the pion channel. It contains the pion pole
which, according to Eq.~\eqref{axcurr}, must be subtracted in order to
obtain the axial TDA.

We can express both, $V(x,\xi,t) $ and $A(x,\xi,t) $, TDAs as the sum of the
active $u$-quark and the active $\bar{d}$-quark distributions. The
first contribution will be proportional to the $d$'s charge, $Q_d$,
and the second contribution to the $u$'s charge, $Q_u$. Therefore, we
can write%
\begin{equation}
D^{\pi^{+}}\left(  x,\xi,t\right)  =Q_d\, d_{u\rightarrow d}^{\pi^{+}
}\left(  x,\xi,t\right)  + Q_u \,d_{\bar{d}\rightarrow\bar{u}}^{\pi^{+}
}\left(  x,\xi,t\right)  \quad,
\label{vtda}
\end{equation}
with $D=V, A$ and $d=v,a$. Isospin relates these two
contributions. For the vector TDA, we obtain $
v_{\bar{d}\rightarrow\bar{u}}^{\pi^{+}}\left( x,\xi,t\right)
=v_{u\rightarrow d}^{\pi^{+}}\left(-x,\xi,t\right) $.
 For the axial TDA we have
$a_{\bar{d}\rightarrow\bar{u}}^{\pi^{+}}(x,\xi,t)=-a_{u\rightarrow
  d}^{\pi^{+}}(-x,\xi,t)$ where the minus sign is originated in the
change in helicity produced by the $\gamma_{5}$ operator.  The $d_{u\rightarrow d}^{\pi^{+}}(x,\xi,t) $ contribution is non-vanishing in the
region $-|\xi|<x<1$ while
$d_{\bar{d}\rightarrow\bar{u}}^{\pi^{+}}(x,\xi,t) $ in the region
$-1<x<|\xi|$. Given the
relation \eqref{vtda}, the support of the whole TDA,
$V^{\pi^{+}}\left( x,\xi,t\right) $ or $A^{\pi^{+}}\left(
  x,\xi,t\right) $, is therefore $x\in[-1,1]$, as required.

\begin{figure}
\begin{minipage}{7cm}
\hspace{-.75cm}
\includegraphics[height=.25\textheight]{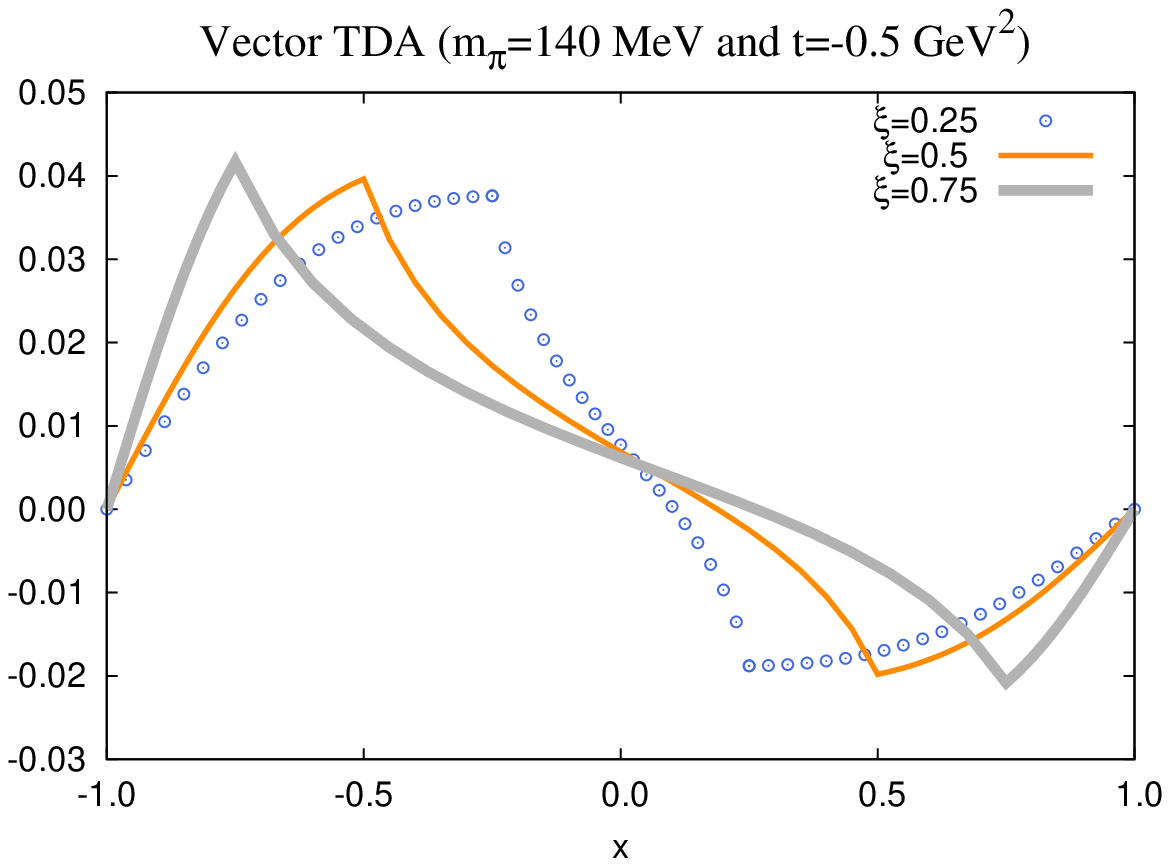}
\end{minipage}
\begin{minipage}{7cm}
\hspace{-.4cm}
\includegraphics[height=.25\textheight]{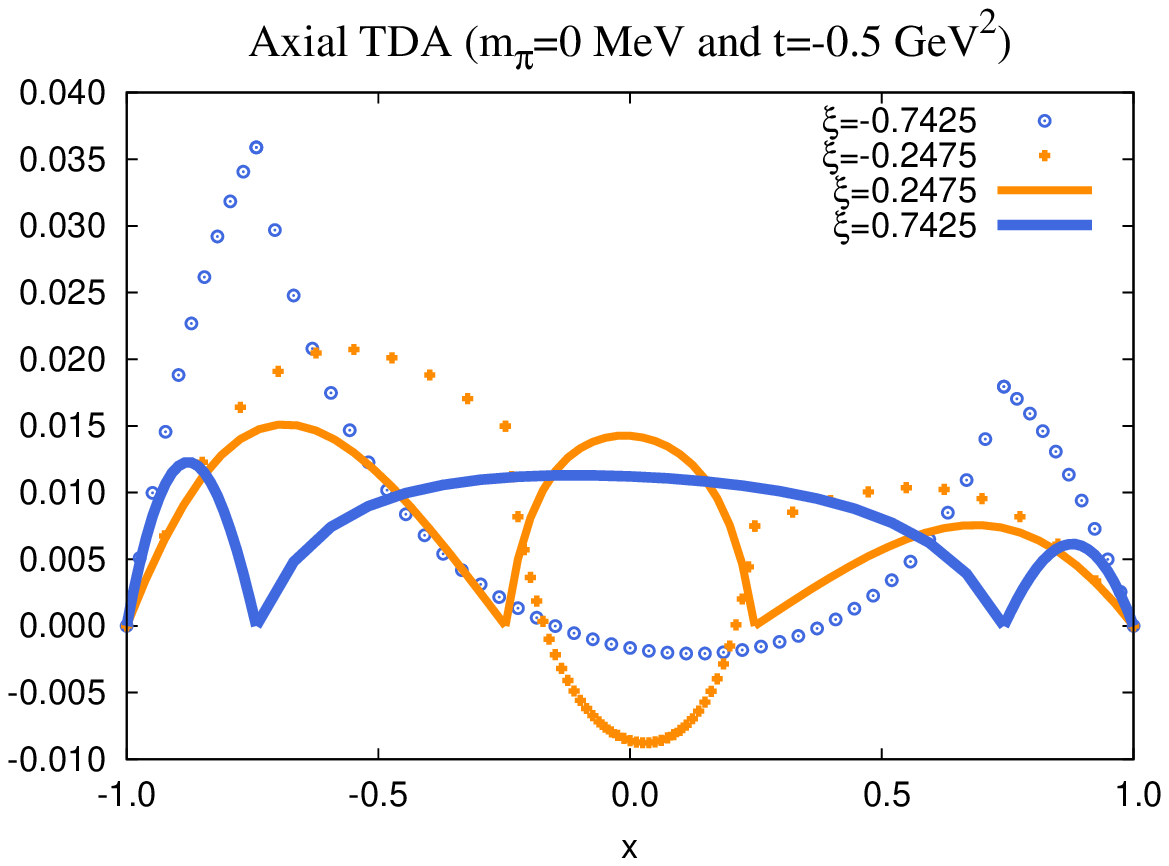}
\end{minipage}
  \caption{Vector  and axial TDA in the NJL model.
The amplitudes are lower for higher $(-t)$ values, as it can be
inferred from the decreasing of the form factors with $\left(  -t\right)  $,
connected to the TDAs through the sum rules.}
\label{vector}
\end{figure}

\begin{figure}
\includegraphics[height=.35\textheight]{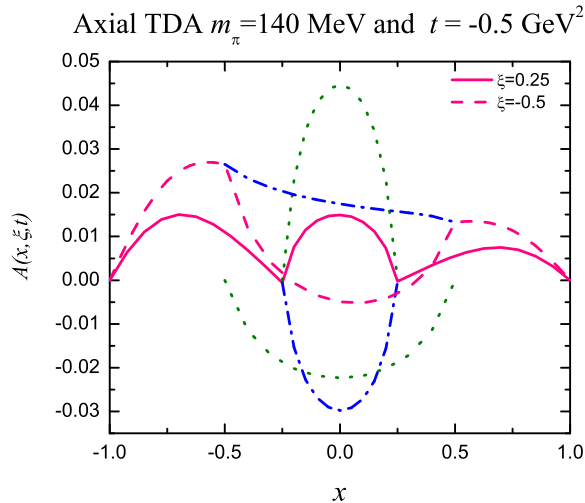}
  \caption{Contributions to the axial TDA
for both positive ($\xi=0.25,$ solid line) and negative ($\xi=-0.5,$ dashed
line) values of the skewness variable. In each case, and in the $x\in [-|\xi|,|\xi|]$ region, the contribution coming from
the first diagram of Fig.~\ref{feyn} is represented by the dashed-dotted lines
and the non-resonant part of the second diagram of Fig.~\ref{feyn} is
represented by the dotted lines. }
\label{axpp}
\end{figure}

In Fig.~\ref{vector} we show the vector and axial TDA
calculated in the NJL model\footnote{For numerical predictions, the
  standard values of the parameters given in
  Ref.~\cite{Klevansky:1992qe} are used.} for different values of the momentum
transfer $t$. The mass of the pion, being either $m_{\pi}=0$ MeV in
the chiral limit or $m_{\pi}=140$ MeV, does not significantly
influence the result. A $\xi$-symmetry is observed for $V(x,\xi,t)$ in the chiral limit:
the vector TDA is an even function of the skewness variable so that we
show the results for positive $\xi$ only.  At the contrary, the shape
of the axial TDA depends on the sign of the skewness variable. The two
distinct behaviours are shown in Fig.~\ref{vector}. For positive $\xi$,
the contribution coming from the second diagram of Fig.~\ref{feyn} is
dominant and produces the maxima around $x=0$. For negative values of
the skewness variable, the contribution of both diagrams have opposite
signs, as shown in Fig.~\ref{axpp}.
Given Eq.~\eqref{vtda}, we can say that isospin relates the value of the
 vector and axial TDAs in the $|\xi|<x<1$ and $-1<x<-|\xi|$ regions,
\beq
V(x,\xi,t)  =-\frac{1}{2}V(-x,\xi,t) \quad \&\quad
A(x,\xi,t)  =\frac{1}{2}A(-x,\xi,t), \qquad\left\vert
\xi\right\vert <x<1\quad.
\label{iso}
\eeq
The factor $1/2$ corresponds to the ratio between the charge of the $u$ and $d$ quarks.
We observe in Fig.~\ref{vector} that our TDAs
satisfy these relations. It must be realized that the relation \eqref{iso}
cannot be changed by evolution.

The obtained TDAs should obey the sum rules, as already
mentioned. For the form factors given in the NJL model (see Appendix
of Ref.~\cite{Courtoy:2007vy}), the sum rules 
\beq
\int_{-1}^{1}dx~D^{\pi^{+}}(x,\xi,t)
=\frac{\sqrt{2}f_{\pi}}{m_{\pi}}F_{D}\left( t\right) \quad,
 \label{sumrule}
 \eeq with $D=V,A$, are recovered. In particular we obtain the value
 $F_{V}^{\pi^{+}%
 }(0)=0.0242$ for the vector form factor at $t=0$, which is in
 agreement with the experimental value $F_{V}(0)=0.017\pm0.008$ given
 in \cite{Yao:2006px}. We also obtain $F_{A}^{\pi^{+}}(0)=0.0239$ for
 the axial form factor at $t=0$, which is about twice the value
 $F_{A}^{\pi^{+}}(0)=0.0115\pm0.0005$ given by the PDG
 \cite{Yao:2006px}.

We expect the TDAs to respect the polynomiality condition. However no
constraint from time reversal enforces the polynomials to be even functions of the skewness variable,
i.e. the polynomials include all powers of $\xi$
\begin{equation}
\int_{-1}^{1}dx\,x^{n-1}\,D(x,\xi,t)=\sum_{i=0}^{n-1}\,C^D_{n,i}(t)\,\xi^{i}\quad,
\label{oddeven}
\end{equation}
where $C^D_{n,i}(t)$ are the generalized form factors.  The relation
\eqref{oddeven} has been numerically verified.  In the chiral limit, we have numerically found that
the odd powers in $\xi$ go to zero for the polynomial expansion of the
vector TDA.

Other studies of pion-photon TDAs have already been done
\cite{Tiburzi:2005nj, Broniowski:2007fs, Kotko:2008gy}. In
Refs.~\cite{Tiburzi:2005nj, Broniowski:2007fs}, double distributions
have been used. Therefore  polynomiality is implemented by definition
and cannot be considered a result.  The aim of the author of
Ref.~\cite{Tiburzi:2005nj} is to provide some estimates of the vector
and axial TDAs on the basis of the positivity bounds. In this way we
must compare only the order of magnitude of the obtained amplitudes,
which is similar to ours.  

\begin{figure}
\begin{minipage}{7.5cm}
\hspace{-.5cm}
\includegraphics[height=.30\textheight]{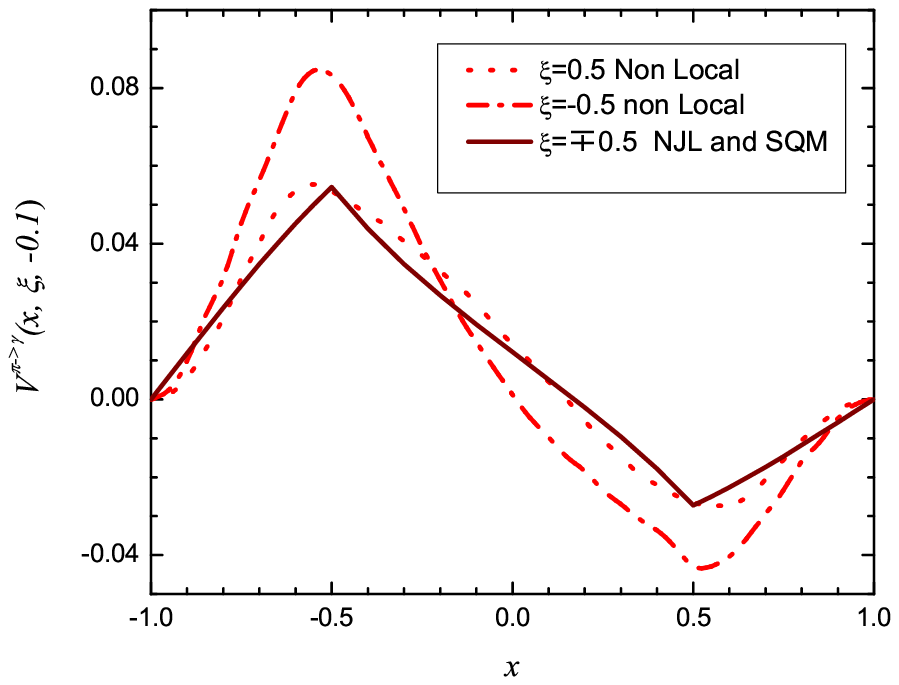}
\end{minipage}
\begin{minipage}{7.5cm}
\hspace{-.5cm}
\includegraphics[height=.31\textheight]{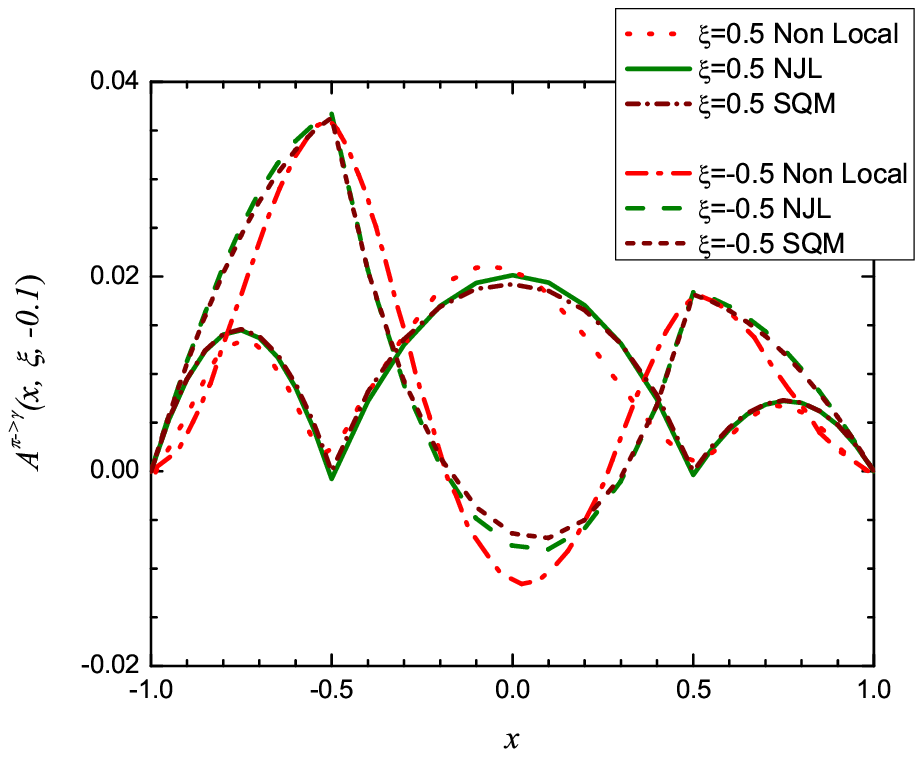}
\end{minipage}
  \caption{ Comparison of the $\pi^+\to\gamma$ TDAs for $t=-0.1$ GeV$^2$ of Refs.~\cite{Broniowski:2007fs, Kotko:2008gy} for $m_{\pi}=0$ MeV  and  Ref.~\cite{Courtoy:2007vy} for $m_{\pi}=140$ MeV.
  On the left, we have: the vector  TDA  for $\xi=\pm 0.5$ as a single (solid) curve  for the results of both  the NJL model and SQM (these four curves are indistinguishable);
    the result of the non-local $\chi$QM calculation  for $\xi=0.5$ (dotted line) and for $\xi=-0.5$ (dashed-dotted line). On the right, we have the results of the axial TDA 
  with the choice for the sign of $\xi$ as discussed in the text.}
 
\label{comp}
\end{figure}

 The vector and axial TDAs calculated in the SQM~\cite{Broniowski:2007fs}, NJL model~\cite{Courtoy:2007vy} and non-local chiral quark model~\cite{Kotko:2008gy}  are compared in Fig.~\ref{comp}.
The authors of the first reference use the so-called asymmetric notation.  The comparison is here awkward since the authors define, after their Eq.~(3), $\zeta=(p_{\gamma}-p_{\pi})^+/p_{\pi}^+$ while, after  their Eq.~(4),  the definition $-\zeta=(p_{\gamma}-p_{\pi}).n$ is given. We nevertheless decide to use  the standard relation between their asymmetric notations and our symmetric ones~\cite{Diehl:2003ny}.
Their functions $V_{\mbox{\tiny SQM}}$ and $A_{\mbox{\tiny SQM}}$  corresponds to our $d^{\pi^+ \to \gamma}$ given in Eq.~\eqref{vtda}.
The normalization condition is different from the one used here and the results quoted by these authors must be corrected, for the vector TDA, by a factor  $48\pi^2\,\sqrt 2 f_{\pi} \,F_V(0)/m_{\pi}\sim 10$ before comparison. From Fig.~\ref{comp} we conclude that there is a
qualitative and quantitative agreement, for the vector TDA, between the results of
Ref.~\cite{Broniowski:2007fs} and those obtained in the NJL model.
Regarding the axial TDA we observe, in addition to the normalization factor $48\pi^2\,\sqrt 2 f_{\pi} \,F_A(0)/m_{\pi}\sim 10$,  a change in the global sign due to different definitions.
In the first version of \cite{Broniowski:2007fs} the axial TDA for  positive values of $\xi$ is given\footnote{There is a typographic error in Eq.~(23) of the first version of this reference, where a factor $M_V^2/6$ must be dropped.}. It coincides with the results in the NJL model,  as observed in Fig.~\ref{comp}.
Surprisingly, the result presented in Ref.~\cite{Broniowski:2007fs} coincides with our result for negative $\xi$ (perhaps due to the change in the definition of $\xi$  mentioned above).

In
Ref.~\cite{Kotko:2008gy} the TDAs are calculated in three different
models. The first one is a local model which pole structure has some similarity with   the one of the  NJL model. The two other models, i.e. semi-local and full non-local,
follow the results of the local one. The most prominent difference
between the results obtained in the non-local and the NJL models is
the appearence of important odd powers in $\xi$ in the polynomial
expansion of the vector TDA.
We know from Ref.~\cite{Noguera:2005ej}
that, for non-local models, there are additional contributions to
those calculated in Ref.~\cite{Kotko:2008gy}. In the case of PDFs,
disregarding these contributions can produce small isospin violations
\cite{Noguera:2005cc}.  It can therefore be considered that, on that point, the results of Ref.~\cite{Kotko:2008gy} must be confirmed.
 For numerical comparison, the results obtained in \cite{Kotko:2008gy}  must be corrected
by a factor $2\pi^2\,\sqrt 2 f_{\pi}\,F_V(0)/m_{\pi}\sim 0.45$ due to the use of a different normalization
condition. 
For the axial TDA, there is a change in the definition of the skewness variable between the caption of Figs.~3, 5 and Fig.~9 of Ref.~\cite{Kotko:2008gy}.
We observe, see  Fig.~\ref{comp},  that our results in the NJL model coincide with the results of the latter reference if the convention of the caption of their  Fig.~9 is chosen.

It can eventually be concluded that there is no disagreement
between the different studies concerning the $\pi$-$\gamma$ TDAs besides the ambiguity in the definition of the skewness variable.
Calculations of Refs.~\cite{Broniowski:2007fs, Kotko:2008gy} are performed in the chiral limit, where $\xi$ runs from $-1$ to $1$.  The symmetric nature of this interval makes difficult to check the sign of $\xi$. On the other hand, the NJL calculation~\cite{Courtoy:2007vy}  is given for physical pion mass. In this case, the kinematics of the process imposes $t/(2m_{\pi}^2-t)<\xi<1$. From Eqs.~(22) and  (23) of Ref.~\cite{Courtoy:2007vy}  we observe that there is a pole in the axial TDA for the limit value $\xi=t/(2m_{\pi}^2-t)$, preventing us from going through unphysical values of $\xi$. Moreover, the sum rule Eq.~\eqref{sumrule} for both the vector and axial TDAs are here satisfied for physical values of $\xi$, and broken in the unphysical regions $\xi<t/(2m_{\pi}^2-t)$ and $\xi>1$. We therefore conclude that the choice of sign in Ref.~\cite{Courtoy:2007vy} is consistent and gives a guideline for  comparing with other models.

The $\pi$-$\gamma$ vector and axial Transition Distribution Amplitudes
have been defined.  The results in different models have been
discussed. In particular, the numerical results of
Ref.~\cite{Broniowski:2007fs, Courtoy:2007vy, Kotko:2008gy} have been compared. The use of a fully
covariant and gauge invariant approach guaranties that all the
fundamental properties of the TDAs will be recovered, in particular,
the right support, i.e. $x\in\left[ -1,1\right] $, the sum rules and
the polynomiality expansions. Hence, in the NJL model, these three
properties are not inputs, but rather results of the
calculation.

Recently, cross section estimates for the process \eqref{gg} have
been proposed in Ref.~\cite{Lansberg:2006fv} using, for the
non-perturbative part, $t$-independent double distributions, in a
first approach, and, in a second, the $t$-dependent results of
Ref.~\cite{Tiburzi:2005nj}.  In the line of sight of this reference,
similar estimates for the meson pair production in hard
$\gamma^{\ast}\gamma$ scattering using the results for the TDAs cited
above could be given \cite{pionpole}. In particular, it would be worth
investigating the model independence of the structure independent
terms \cite{Pire:2007ut} and then estimating the pion pole
contribution to the cross section \cite{pionpole}.

\begin{theacknowledgments}
This work has been supported  by the Sixth Framework Program of the
European Commision under the Contract No. 506078 (I3 Hadron Physics); 
by the MEC (Spain) under the Contract FPA 2007-65748-C02-01 and 
the grant AP2005-5331 and by EU FEDER. Feynman diagrams drawn using JaxoDraw \cite{Binosi:2003yf}.
\end{theacknowledgments}

\end{document}